\documentclass[aip,jcp,amsmath,amssymb,floatfix,reprint,citeautoscript]{revtex4-1}

\usepackage[utf8]{inputenc}
\usepackage{ae}
\usepackage{graphicx}
\usepackage{bibunits}
\usepackage[version=4]{mhchem}
\usepackage{wrapfig}
\usepackage[colorlinks,allcolors=black,citecolor=blue,urlcolor=blue]{hyperref}

\newcommand{\dd}{\mathrm{d}}

\defaultbibliography{paper,revtex-control}
\defaultbibliographystyle{aipnum4-1}

\graphicspath{{./Figures/}}

\begin{document}

\def\mytitle{Benzene radical anion in the context of the Birch reduction: when~solvation~is~the~key}
\title{\mytitle}

\author{Krystof Brezina}
\affiliation{
Charles University, Faculty of Mathematics and Physics, Ke Karlovu 3, 121 16 Prague 2, Czech Republic
}
\affiliation{
Institute of Organic Chemistry and Biochemistry of the Czech Academy of Sciences, Flemingovo nám. 2, 166 10 Prague 6, Czech Republic
}

\author{Pavel Jungwirth}
\affiliation{
Institute of Organic Chemistry and Biochemistry of the Czech Academy of Sciences, Flemingovo nám. 2, 166 10 Prague 6, Czech Republic
}

\author{Ondrej Marsalek}
\email{ondrej.marsalek@mff.cuni.cz}
\affiliation{
Charles University, Faculty of Mathematics and Physics, Ke Karlovu 3, 121 16 Prague 2, Czech Republic
}

\date{\today}

\begin{abstract}

\setlength\intextsep{0pt}
\begin{wrapfigure}{r}{0.35\textwidth}
  \hspace{-1.8cm}
  \includegraphics[width=0.35\textwidth]{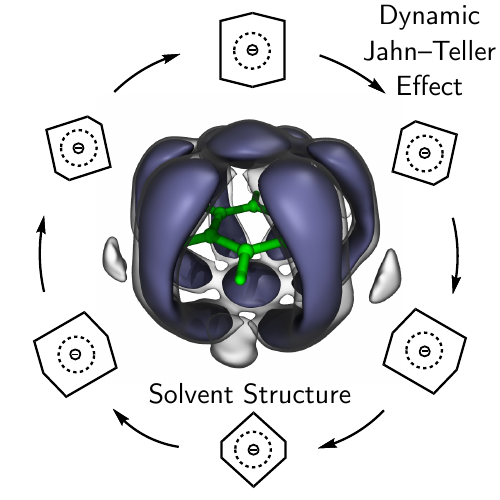}
\end{wrapfigure}

The benzene radical anion is an important intermediate in the Birch reduction of benzene by solvated electrons in liquid ammonia.
Beyond organic chemistry, it is an intriguing subject of spectroscopic and theoretical studies due to its rich structural and dynamical behavior.
In the gas phase, the species appears as a metastable shape resonance, while in the condensed phase it remains stable.
Here, we approach the system by \textit{ab initio} molecular dynamics in liquid ammonia and demonstrate that the inclusion of solvent is crucial and indeed leads to stability.
Beyond the mere existence of the radical anion species, our simulations explore its condensed-phase behavior at the molecular level and offer new insights into its properties.
These include the dynamic Jahn--Teller distortions, vibrational spectra in liquid ammonia, and the structure of the solvent shell, including the motif of a $\pi$-hydrogen bond between ammonia molecules and the aromatic ring.

\end{abstract}

{\maketitle}

\begin{bibunit}

\nocite{revtex-control}


One of the most prominent roles of the benzene radical anion (\ce{C6H6^{.-}}) in organic chemistry is that of a reactive intermediate in the Birch process~\cite{Birch1944,Birch1945,Birch1946,Birch1947a, Birch1947,Birch1949} (Equation \ref{eq:birch}) used to reduce benzene or related derivatives into 1,4-cyclohexadienes.
Such reduction is realized through the action of excess electrons.
Specifically, the reaction~\cite{Clayden2012}
\begin{equation}
    \includegraphics[width=0.90\linewidth]{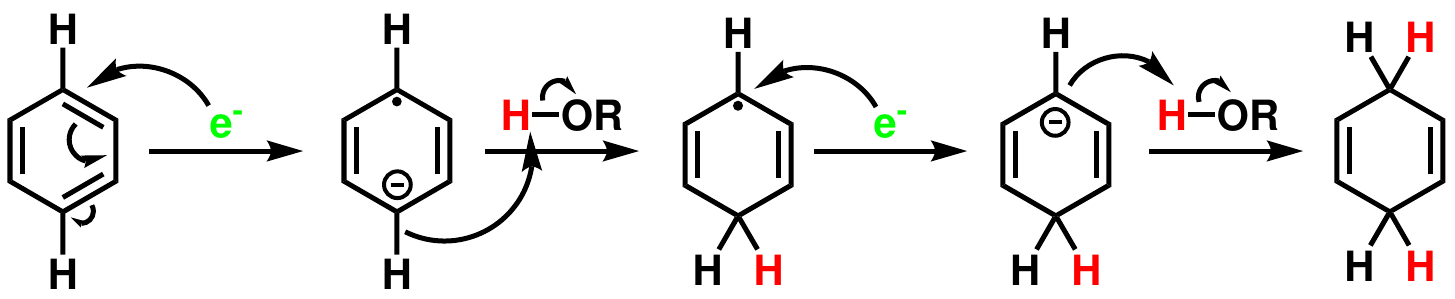}
    \label{eq:birch}
\end{equation}
is initiated by introducing the aromatic substrate together with a proton source (usually a simple alcohol) into the dark blue solution of solvated electrons~\cite{Buttersack2020/10.1126/science.aaz7607} in liquid ammonia~\cite{Boese2003/10.1063/1.1599338,Buttersack2019}, created by dissolution of an alkali metal~\cite{Zurek2009}.
Already in the reaction's first step, the benzene radical anion is formed when a neutral benzene molecule accepts an excess electron from the surrounding solution, allowing the substrate to enter the reduction pathway.
The Birch reduction has proved to be an important tool for selective reduction of organic substrates both industrially and in organic synthesis~\cite{Birch1992,Birch1996,Joshi2005}.
However, interest in the anion reaches beyond organic chemistry.
Its open-shell electronic structure defined by a highly symmetric molecular geometry gives rise to phenomena that have been the subject of numerous theoretical and spectroscopic studies since the 1950s~\cite{Hinde1978,Tuttle1958,Shida1973,Moore1981,Birch1980}.
Computationally, the anion presents a challenge due to its need for both a large basis set and a sufficient level of electronic structure theory to accurately capture its properties in the gas phase.
Recently, Bazante and co-workers~\cite{Bazante2015} have published a detailed gas-phase \textit{ab initio} study focusing on the isolated species, specifically its molecular geometry, including the Jahn--Teller (JT) distortions, electron paramagnetic resonance (EPR) spectroscopy parameters, and on confirming its previously suggested~\cite{Hinde1978} character of a metastable shape resonance in vacuum.
Experimentally, the character of the gas-phase benzene radical anion as a metastable resonance was studied using electron transmission spectroscopy~\cite{Sanche1973/10.1063/1.1679228,Jordan1976/10.1021/ja00421a058} where the results even determine the extremely short lifetime of the species on the order of 10~fs.
In contrast, the species has been observed in the condensed phase using multiple experimental approaches.
This includes spectroscopic studies of film-like co-deposits of an alkali metal with benzene using vibrational spectroscopy~\cite{Moore1981} and in the presence of a polar organic solvent using ESR~\cite{Tuttle1958,Hasegawa1994}, or electronic spectroscopy~\cite{Shida1973}.
The issue of electronic stability in solvated environments was directly addressed experimentally in the context of small aqueous clusters~\cite{Maeyama1997/10.1021/jp9721661} and the thermodynamic equilibrium with solvated electrons was measured at various temperatures in a tetrahydrofuran solution~\cite{Marasas2003/10.1021/jp026893u}.
These studies point to the presence of a radical species and suggest that its gas-phase metastability is removed in in the condensed phase: a stable bound state is observed.
Electrochemical properties and the reactivity of the radical anion in solution have also been studied by cyclic voltammetry in dimethoxyethane with a proton source~\cite{Mortensen1984/10.1002/anie.198400841}.
The condensed-phase-induced stability is also implied by the high experimental yields~\cite{Birch1944,Wilds1953,Hook1986,Brandsma1990} of the Birch reduction that would not be achievable in case an intermediate was electronically unstable.
This experimental evidence shows that this stabilization is not solvent specific but rather general to a broad class of polar solvents, implying that the stabilizing effect of the solvent is due to the dielectric environment that it creates around the anion.
It is therefore clear that the solvent environment plays a crucial role in the stabilization of the benzene radical anion in solution.

Our aim in this work is to shed new light on this species using a computational approach.
To this end, we use \textit{ab initio} molecular dynamics (AIMD) to gain insight into and understand in detail the static and dynamic properties of the benzene radical anion in solution, in contrast with previous gas-phase studies, because experimentally the inclusion of the solvent appears to be the key
to obtaining an electronically and thermodynamically stable species.
Specifically, we perform our condensed-phase simulations in explicit liquid ammonia, as this is the natural environment for excess electrons in the context of the Birch reduction.
Using these simulations, we characterize the JT distortions of the radical anion, explore the solvent structure of liquid ammonia around the anionic solute, and predict its contribution to vibrational spectra, compared to those of its neutral counterpart in all cases.


\begin{figure}[tb!]
    \centering
    \includegraphics[width=\linewidth]{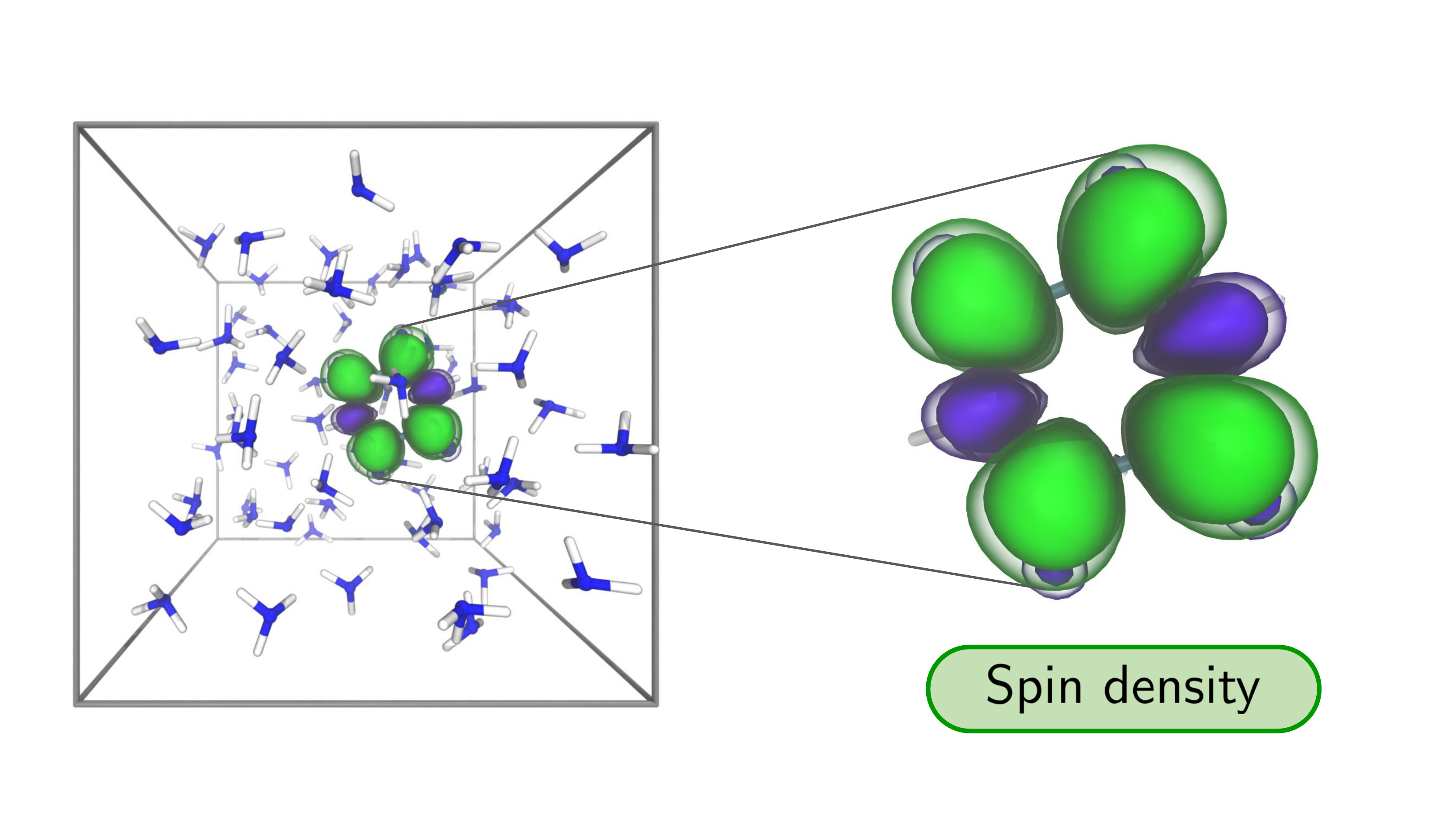}
    \caption{
    Representative snapshot from the hybrid DFT AIMD simulation.
    On the left, we show a snapshot of the whole simulation box containing one benzene, 64 ammonia molecules, and an excess electron with the spin density of the system colored green (positive part) and violet (negative part).
    The close-up shows a detail of the localization of the spin density distribution over the benzene radical anion.
    }
    \label{fig:sim-shapshot}
\end{figure}

The subsequent discussion of our results is based on AIMD simulations of the benzene radical anion and neutral benzene in liquid ammonia at 223~K under periodic boundary conditions using a hybrid density functional theory (DFT) electronic structure calculation on the fly.
AIMD methodology at the hybrid DFT level is required to obtain an electronically stable anionic species with most of the excess electron density localized on the aromatic ring [around 95~\% as determined by the Hirshfeld population analysis\cite{Hirshfeld1977} (see Figures~\ref{fig:spin_quant} and~\ref{fig:spin_qual})], as expected for an electron occupying a bound state in solution.
This is necessary only due to the presence of the anionic radical species: liquid ammonia itself is already well described at lower levels of DFT theory~\cite{Diraison1999/10.1063/1.479194}.
A detailed discussion of the employed methodology~\cite{Vandevondele2005/10.1016/j.cpc.2004.12.014,Hutter2014/10.1002/wcms.1159} is presented in the Supporting Information; here, we show a snapshot illustrating the localization of the excess electron density (green and violet contours of the spin density) on the aromatic ring in Figure~\ref{fig:sim-shapshot} and a trajectory video file (detailed in Section~\ref{sec:visualizations}) that shows also the time evolution of the system, including the spin density.


\begin{figure}[tb!]
    \centering
    \includegraphics[width=0.6\linewidth]{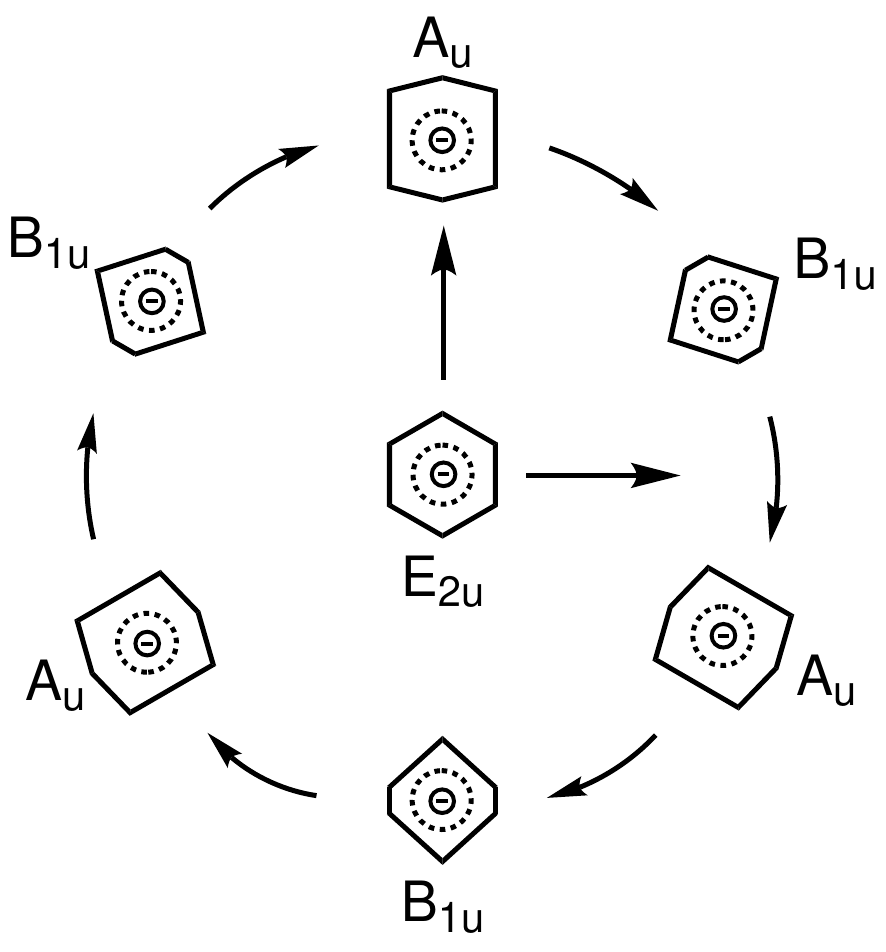}
    \caption{
    The JT pseudorotation of the benzene radical anion.
    The central hexagonal structure distorts to attain one of the lower-symmetry and lower-energy structures around the pseudorotational path around which it then circles.
    }
    \label{fig:pseudorotation}
\end{figure}

With regard to the molecular structure of the radical anion, quantum theory predicts that a symmetry breaking of the initially perfectly hexagonal molecule of benzene is expected via a JT distortion after the addition of the excess electron in a degenerate lowest unoccupied electronic state.
More precisely, the initial $\mathrm{E_{2u}}$ electronic state in the $D_{\mathrm{6\mathrm{h}}}$ point group distorts to two possible distinct states labeled $\mathrm{A_{u}}$ and $\mathrm{B_{1u}}$ in the $D_{\mathrm{2\mathrm{h}}}$ nearest lower symmetry point group.
A geometric relaxation of the molecular structure follows, resulting in two possible shapes that can be roughly described as contracted ($\mathrm{B_{1u}}$) and elongated ($\mathrm{A_{u}}$) hexagons that have been previously predicted by gas-phase calculations~\cite{Hinde1978,Bazante2015} and experimentally~\cite{Hasegawa1994}.
For comparison, the related but computationally simpler benzene radical cation, which has almost identical distortions, does exist as a stable species already in the gas phase and has been studied by AIMD before~\cite{Tachikawa2018}.
The energy barriers between the two JT structures of the radical anion are expected to be low enough to allow for a dynamic JT effect where the structure is not rigid but undergoes a pseudorotation~\cite{Bazante2015} (Figure~\ref{fig:pseudorotation} and the video file detailed in Section~\ref{sec:visualizations}).
Below, we analyze the structural distortions of the anion as observed in our condensed-phase simulations and discuss whether and how the JT behavior manifests in solution.

\begin{figure}[b!]
    \centering
    \includegraphics[width=\linewidth]{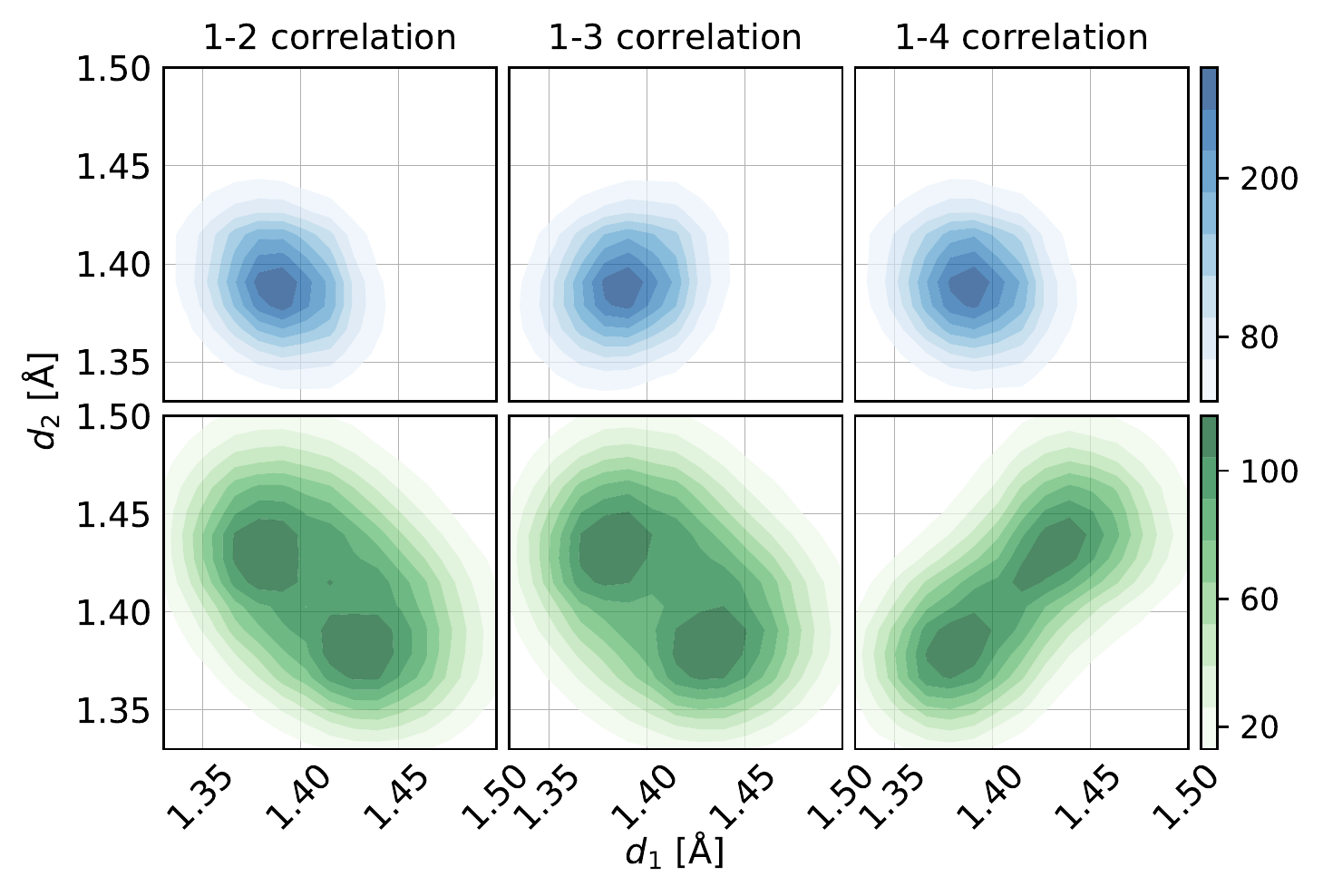}
    \caption{
    Correlations of carbon--carbon bond lengths ($d$) between bonds of different neighboring schemes.
    The uncharged system data can be found in the top row (blue), whereas the bottom row contains the anionic data (green).
    }
    \label{fig:JT1}
\end{figure}

Initially, we analyze the structure of the solute by correlating pairs of carbon--carbon bond lengths around the aromatic ring. 
Specifically, we first correlate all pairs of directly neighboring bonds (1-2 correlation), for all bond pairs with one additional bond in between (1-3 correlation), and finally all bond pairs neighboring over two extra bonds (1-4 correlation).
The resulting bivariate probability densities are shown in Figure~\ref{fig:JT1} for both neutral benzene (top row, blue) and benzene radical anion (bottom row, green).

The neutral benzene reference data shows unimodal distributions in all three panels positioned on the main diagonal.
This implies a hexagonal symmetry of the neutral solute and together with the mean carbon--carbon bond length of 1.39~\AA\ suggests that its structure is very similar to that in the gas phase: this is expected as no symmetry-lowering distortions are anticipated at this point.
In contrast, adding the excess electron to the solute changes these results dramatically.
Bimodal distributions appear, and moreover, the 1-2 and 1-3 correlations now clearly show a negative correlation; on the contrary, the remaining 1-4 correlation displays a positive one.
These observations are consistent with the predicted $\mathrm{A_{u}}$ and $\mathrm{B_{1u}}$ JT structures that require a bond length heterogeneity in the 1-2 and 1-3 pairs (\textit{i.e.}, having a shorter and a longer bond in the pair, respectively) and a homogeneity in the 1-4 pair where the bond lengths have to match due to the presence of a center of symmetry in the $D_{2\mathrm{h}}$ point group. 

\begin{figure}[t!]
    \centering
    \includegraphics[width=\linewidth]{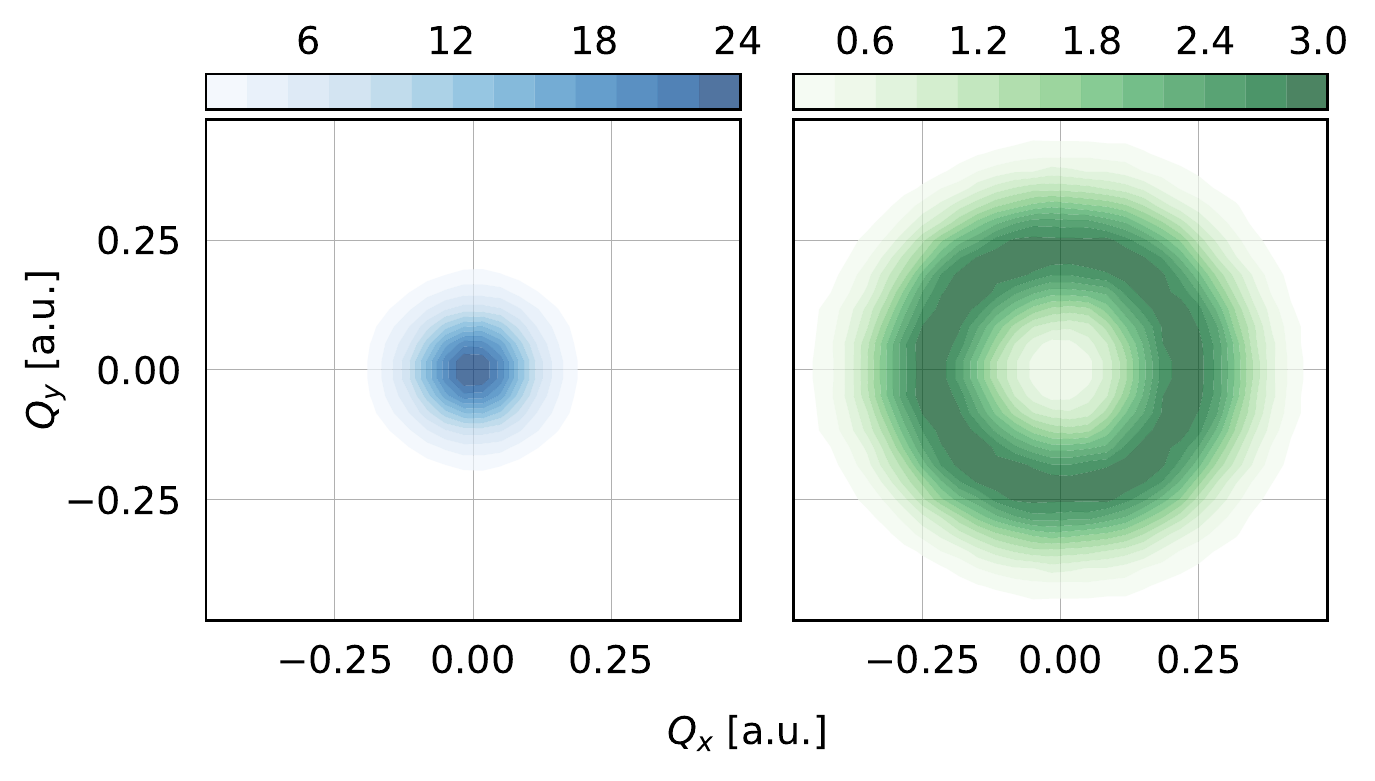}
    \caption{
    Correlations of carbon ring-deforming JT-active normal mode amplitudes for the neutral (left, blue) and anionic (right, green) systems.
    }
    \label{fig:JT2}
\end{figure}

The analysis presented above using bond length correlations unarguably shows the presence of symmetry-lowering molecular distortions of the anion that agree with the previously predicted JT structures.
However, as it averages over \textit{all} bond pairs around the ring, it is in fact insensitive to distinguishing the individual \textit{pseudorotamers}.
Moreover, the bond lengths do not provide a meaningful reaction coordinate for evaluating the energetics of the distortions.
Therefore, we present a complementary analysis where we project the immediate distortions of the solute onto the harmonic vibrational normal modes of the neutral benzene gas-phase reference and observe the correlations of degenerate pairs of modes of the JT active $e_{\mathrm{2g}}$ symmetry.
Only one of the four possible JT active normal mode pairs yields a correlation that rises above thermal fluctuations, and thus, we track the JT distortions using this particular pair of modes, averaging over all remaining degrees of freedom (Figure~\ref{fig:JT2}).
The pair in question describes the vertical elongation of the carbon ring and its sideways-skewing motion~\cite{Preuss2006}, is located at the frequency of 1654~cm$^{-1}$ in the gas phase, and will be denoted further as $Q_x$ and $Q_y$.
Linear combinations of these two modes can reconstruct the crucial features of the distortions sketched in Figure~\ref{fig:pseudorotation}.
A video file detailed in Section~\ref{sec:visualizations} visualizes the propagation of the benzene ring along a circular trajectory in the subspace spanned by these two modes.  

Once again, a trivial, origin-centered peak appears in the neutral system which corresponds to a symmetric structure with respect to the two studied modes.
Its narrowness implies a certain level of structural rigidity.
In contrast, the anion displays a typical ``Mexican hat'' shape with the area around the center being depopulated and the main population residing in a ridge around it.
This ridge directly corresponds to the pseudorotational path sketched out in Figure~\ref{fig:pseudorotation}.
Such an observation is in agreement with the prediction of the JT effect and indeed shows that the structure of the radical anion is not rigid, but rather flexible and dynamic.
As we discuss in the Supporting Information and show in Figure~\ref{fig:vacf}, the JT distortion changes on a very short time scale of no more than $100$~fs.
While the symmetric structure with respect to the two studied modes is penalized by a free energy barrier of around 2~kT, the free energy variation along the pseudorotational path turns out to be small relative to thermal fluctuations, as detailed in Figure~\ref{fig:1D-profiles} and the corresponding discussion in the Supporting Information.

\begin{figure}[tb!]
    \centering
    \includegraphics[width=\linewidth]{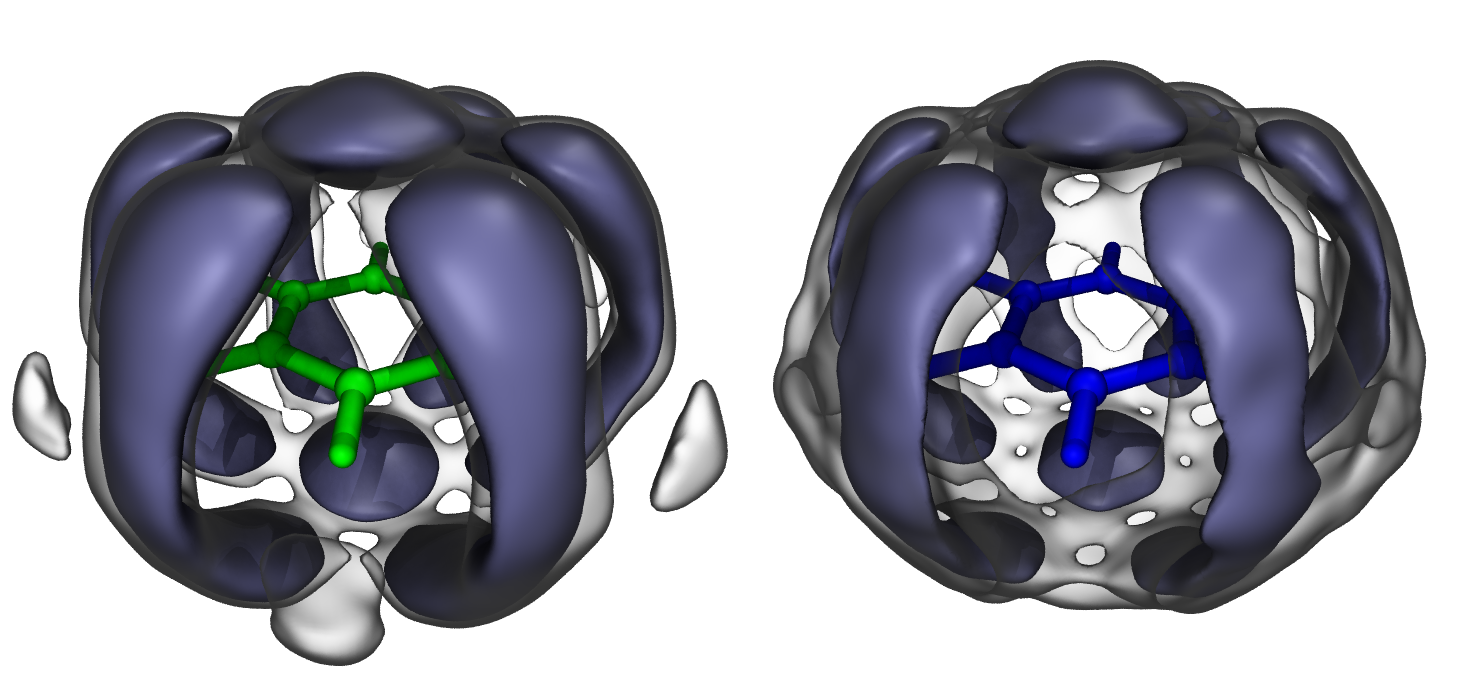}
    \caption{
    Spatial distribution functions of ammonia nitrogen atoms around the center of mass of the aromatic ring.
    Anionic data is shown on the left (green solute), and neutral data on the right (blue solute).
    The blue--purple contour shows regions where the density is 4.5 times that of a non-interacting solvent.
    For the purpose of presentation, the three-dimensional histograms were convolved with a $\sigma = 1\:\text{\AA}$ Gaussian filter.}
    \label{fig:sdfs}
\end{figure}

Having addressed the features of the intrinsic structure of the solute, we now turn our attention to the structure of the solvent.
The specific solvation structure is captured in the nitrogen spatial distribution functions (SDFs) around each of the two (anionic and neutral) solutes shown in Figure~\ref{fig:sdfs}.
The SDFs take on a cagelike form with maxima in strips around the carbon--carbon midbond regions and in the areas above and under the ring.
Intuitively, these areas, clearly separated from the rest of the solvation patterns, can be interpreted as a manifestation of the $\pi$-hydrogen bonding phenomenon, as discussed below.
The SDFs of both species look similar to the naked eye: the only visible difference is a more blurred character in the neutral benzene case, suggesting that the diffuse negative charge on the anion does provide a structuring effect on the solvent arrangement, but not a very substantial one.
The JT distortions of the anion are not expected to alter the symmetry of the solvation structure: the distortions happen symmetrically in three different directions, which would lead to a cancellation of their effect on the shape of the SDF in the long time limit.
Moreover, the solvent does not have enough time to rearrange and accommodate the immediate JT distortions of the solute anyway due to their previously mentioned very short time scale.

\begin{figure}[b!]
    \centering
    \includegraphics[width=\linewidth]{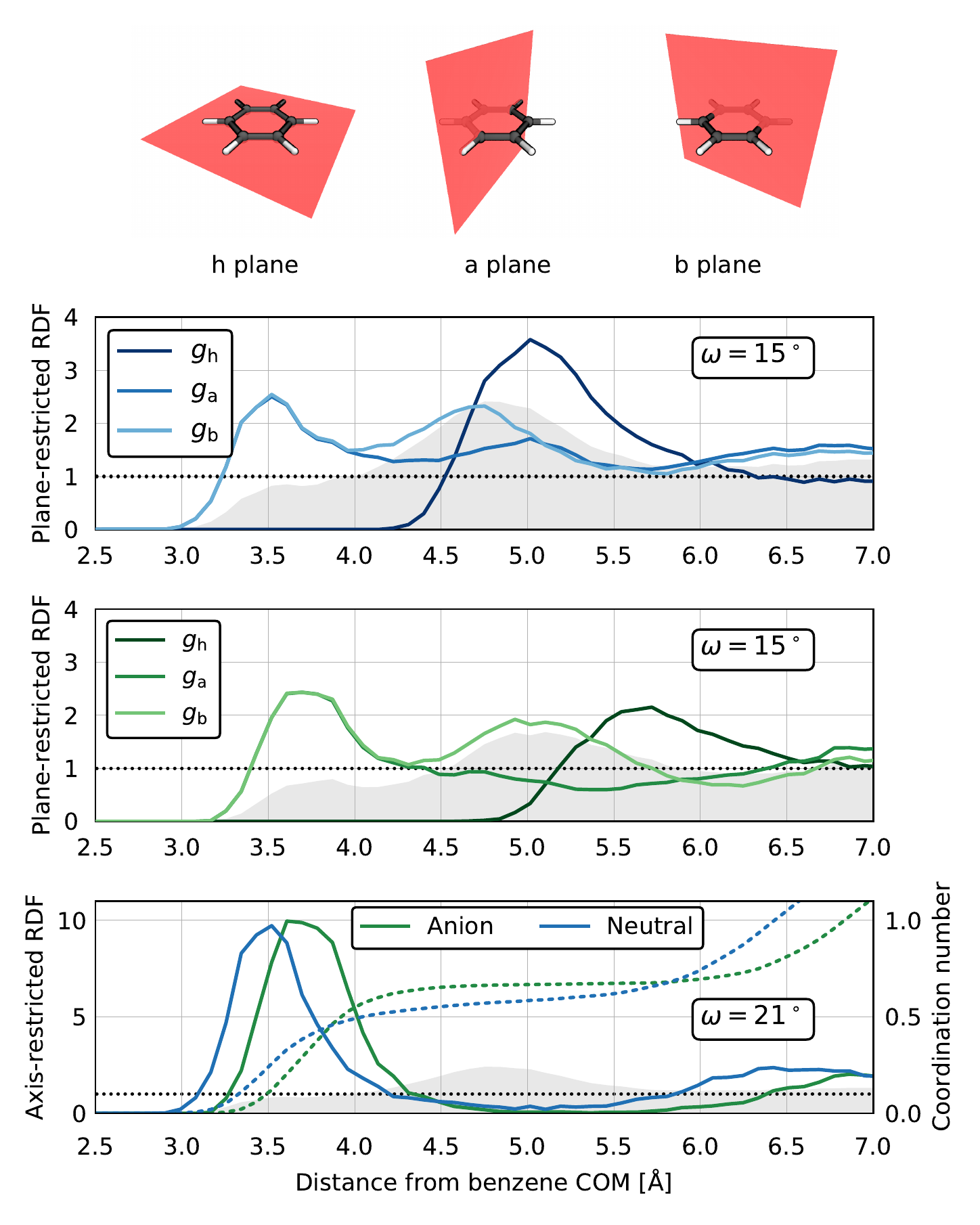}
    \caption{
    RRDFs of ammonia nitrogen atoms around the COM of the aromatic ring.
    RRDFs restricted to significant planes of the solute geometry (top row) under a restriction angle of $\omega = 15^{\circ}$ are shown in the top (neutral benzene) and middle (radical anion) panels.
    The gray shading represents the unrestricted RDF ($\omega = 90^{\circ}$) in both cases.
    In the bottom panel, axis RRDFs focus on the $\pi$-hydrogen cloud with a restriction angle of $\omega = 21^{\circ}$ for the radical anion (green) and neutral benzene (blue) with the gray shaded neutral benzene unrestricted RDF for size comparison.
    Dashed lines show the per-site running coordination numbers of the axis RRDFs.
    }
    \label{fig:rdfs}
\end{figure}

In Figure~\ref{fig:rdfs}, we plot the plane-restricted radial distribution functions (RRDFs) of ammonia nitrogen atoms around the center of mass (COM) of both the neutral and anionic solutes, obtained formally by partial angular integration of their respective SDFs (see the Supporting Information for the specifics of the procedure and Figure~\ref{fig:angle} for its convergence).
These RRDFs describe the radial solvent distribution close to certain given planes instead of the whole sphere (as in the usual plain RDF).
Therefore, they provide more detailed information about the anisotropic solvation environment than a plain unrestricted RDF does.
In total, there are three categories of perpendicular significant planes with respect to the geometry of the studied solutes: the horizontal (h) plane in which the aromatic ring of ideal geometry would lie, three equivalent vertical planes (a) perpendicular to h cutting through a pair of distal carbon atoms, and three more equivalent vertical planes (b) perpendicular to both h and a and cutting through two opposing carbon--carbon bonds (Figure~\ref{fig:rdfs}, top row).
A comparison between the RRDFs in both presented panels further quantitatively confirms the previous qualitative assessment based on the SDFs that the solvent structure around the anionic and the neutral structures is, despite minor differences, similar.
We shall discuss mainly the anionic data (Figure~\ref{fig:rdfs}, middle panel) while keeping in mind that a similar discussion applies to neutral benzene.
In the $g_{\mathrm{h}}$ RRDF (dark green), which addresses the shape of the solvation structure around the horizontal plane, one can observe a rather late onset due to steric shielding by the solute itself followed by a single peak.
The maximum of the first solvation shell of the benzene anion in the horizontal direction is located approximately 5.6~\AA\ (5.0~\AA\ for benzene) from the center of mass.
The $g_{\mathrm{a}}$ RRDF (medium shade of green) shows a similar shape, even though the onset is at a much shorter distance in this case as there is no steric shielding by the solute in the vertical direction.
The observed peak at 3.7~\AA\ (3.5~\AA\ for benzene) corresponds to the $\pi$-hydrogen cloud and is followed by a wide shallow minimum as there is no significant density of solvent around the carbon atoms.
In contrast, the $g_{\mathrm{b}}$ RRDF (light green) shows a second additional peak at 5.0~\AA\ (4.7~\AA\ for benzene) arising from the SDF areas not belonging to the $\pi$-hydrogen cloud.
Both peaks of the RRDF thus belong to the first solvation shell.
This splitting slightly transfers to the $g_{\mathrm{a}}$ plane of neutral benzene (medium blue), which can be considered a sign of its more diffuse (weaker) solvation structure in comparison to that of the anion.
However, we note in passing that the solvation cavity of the anion is systematically larger in all directions.
The onset of the second solvation shell can be noticed in the vertical RRDFs above 6~\AA but is cut short by the employed simulation box size of approximately 14~\AA.
The second solvation shell is not captured in the horizontal RRDF at all.

To close the discussion, we turn briefly to the phenomenon of $\pi$-hydrogen bonding, \textit{i.e.}, bonding between the polar solvent and the $\pi$-electron cloud of the solute.
This behavior has been found in aqueous solutions of benzene both experimentally~\cite{Gierszal2011} and computationally~\cite{Allesch2007}.
In our liquid ammonia simulations we observe a well separated peak in both SDFs, symmetrically located above and under the h plane, indicative of the presence of a $\pi$-hydrogen bond.
Visual inspection of the solvent molecules bound at the site reveals solvent orientation consistent with $\pi$-hydrogen bonding.
These occurrences are found to have a temporary character: molecules do bind and unbind on the simulated time scales.
Once again, we use the procedure of RDF restriction to quantify the $\pi$-hydrogen cloud of the SDFs: this time we restrict the RDF to an axis in a manner complementary to the planar restrictions above (again, see the Supporting Information for additional details).
A natural choice of this axis is the normal of the benzene horizontal plane, \textit{i.e.}, its $C_6$ axis.
In this case, the restriction angle was set to $\omega = 21^{\circ}$ purely on the basis of the SDF geometry in order to completely isolate the whole $\pi$-hydrogen cloud from the rest of the SDF.
The obtained RRDFs can be found in the bottom panel of Figure~\ref{fig:rdfs} for both the anionic solute and the neutral solute.
The $\pi$-hydrogen peak here is massive, reaching a magnitude of 9--10 (\textit{cf.} the gray shaded unrestricted RDF reference).
In the same panel, we also present the corresponding coordination numbers to quantify the solvent population of one of the two $\pi$-hydrogen binding sites located above and under the planar solute.
Even here, it appears that the two studied solutes behave rather similarly and the excess charge on the anion only slightly enhances the interaction: the specific populations are $\sim$0.65 and $\sim$0.70 molecule per binding site for the neutral and anionic solutes, respectively.

\begin{figure}[t!]
    \centering
    \includegraphics[width=\linewidth]{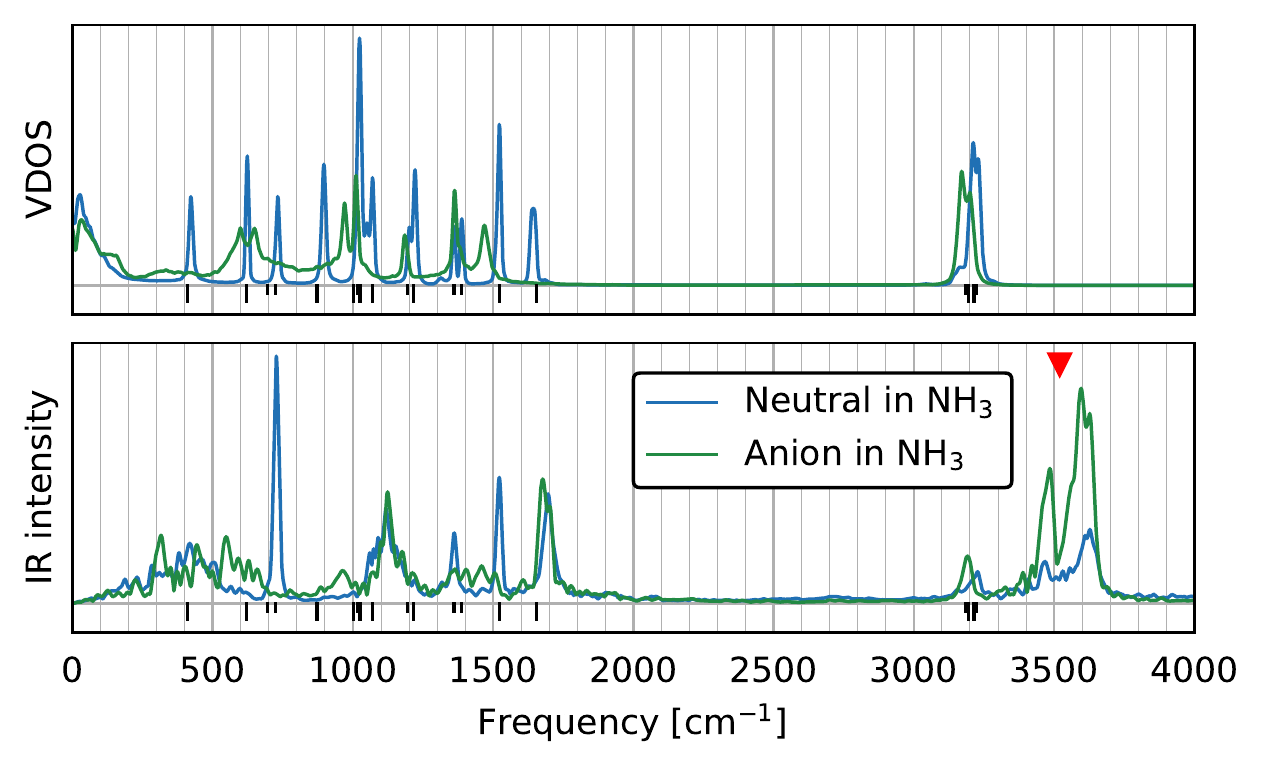}
    \caption{
    Solute-associated vibrational spectra of neutral benzene (blue) and benzene radical anion (green) in liquid ammonia.
    The top panel shows the VDOS, while the bottom panel shows the IR intensity.
    Gas-phase calculated harmonic vibrational frequencies of neutral benzene are shown in the form of black sticks (doubled in length for degenerate modes).
    The red triangle points to the group of peaks discussed in the text.
    }
    \label{fig:vib}
\end{figure}

To be able to relate to spectroscopic measurements, we predict the vibrational spectra of both species in liquid ammonia in the form of the vibrational density of states (VDOS) and infrared intensity (IR).
Theoretical calculations of vibrational spectra allow, unlike their experimental measurement, for a division of the overall spectrum into atomic (or molecular) contributions.
Figure~\ref{fig:vib} shows, for both systems, the atomic VDOS of the solute and the solute-associated IR spectrum, \textit{i.e.}, only terms including a contribution by the solute dipole.
Methodological details and a further decomposition of the IR spectrum into the solute-only and solute-solvent terms can be found in Figure~\ref{fig:cross-term}.
In comparison to its harmonic gas-phase frequencies (stick spectrum in Figure~\ref{fig:vib}), neutral benzene in solution manifests vibrational patterns that can be clearly related to this reference by peak positions.
For the anion, this remains obvious only in the well-separated C-H stretch region ($\sim$3200~cm$^{-1}$).
In the aromatic fingerprint region, the JT distortions nontrivially alter the spectrum; substantial peak broadening and shifts mean that peak assignment is no longer straightforward.
A possible way to identify the anionic species in IR could be based on comparing the regions around 700 and 1500~cm$^{-1}$, where the pronounced peaks of the neutral species are no longer present in the spectrum of the anion.
Consistent with experimental findings in metal--benzene films~\cite{Moore1981}, we observe a slight red shifting in a majority of the anionic peaks that can be related to their neutral counterparts.
Interestingly, we observe additional peaks in the IR that are not present in the VDOS of the solute.
Namely, this is the group of peaks around 3500~cm$^{-1}$ whose presence suggests that a dipole response exists in the solute without a corresponding change in its atomic positions.
These peaks likely arise as a consequence of the perturbation of the electronic structure of the solute caused by the presence of $\pi$-hydrogen bonds.
This is supported by several facts such as the high frequency of the vibration comparable to the N-H stretch in ammonia~\cite{Bromberg1977}, the fact that isolated benzene contains no normal modes at the corresponding frequencies, the fact that the solute--solvent cross-term spectrum (see the Supporting Information) shows a contribution to these peaks, and the fact that benzene solvated in nonpolar environments (such as liquid benzene itself) does not show these additional peaks in the IR spectra~\cite{Thomas2015}.

The presented condensed-phase AIMD simulations bridge the gap between experimental observations and gas-phase calculations of the benzene radical anion, while providing crucial insights into the structure and dynamics of the species at the molecular level.
Using a combination of advanced computational methodology and an explicit representation of the solvent environment, we were able to simulate a radical anion that is electronically stable in solution over extended periods of time, in agreement with its real-world behavior.
Our observation of the localization of the spin density over the aromatic ring points to the existence of a bound state.
Explicit confirmation that the excess electron on the radical anion is bound with respect to the vacuum level is a possible subject of a follow-up study that would build on top of the simulations presented here.
Although our results do not directly address the thermodynamic stability of the benzene radical anion relative to solvated electrons,
the species persists in our simulations for tens of picoseconds, enough to be characterized in a meaningful way.
Furthermore, its thermodynamic stability is implied, at least in the sense of equilibrium with solvated electrons, by the fact that the radical anion is experimentally measurable.
The most prominent structural feature of our condensed-phase benzene radical anion is the presence of JT distortions that are structurally consistent with previous experimental studies and theoretical studies of the gas-phase metastable species in finite basis sets.
We also identified clear differences between the vibrational spectra of the radical anion and its neutral counterpart.
Because of the explicit representation of the solvent environment, our results also offer information about the structure and dynamics of the solvation shell.
A particularly intriguing aspect of the behavior of the solvent arises in the form of $\pi$-hydrogen bonding.
This interaction appears clearly in our solvent distributions and leaves a specific imprint in the IR spectrum associated with the solute, enabled by the dipole response of the solute to the vibrational motion of the solvent.
Our simulations do not contain the proton source, such as an aliphatic alcohol, required for the Birch reduction to proceed.
We can, however, hypothesize that should it be introduced,
its participation in this $\pi$-hydrogen bond might constitute the initial protonation step of the Birch reduction mechanism sketched schematically in Equation~\ref{eq:birch}.
Thus, the demonstrated ability of AIMD to treat this solute has the potential to bring us closer to explicit modeling of the chemical reactivity in the Birch reduction as a whole.
Such a study would provide a complement to the treatment of reaction mechanisms in organic chemistry that is often empirical and simplified, as illustrated, for instance, by the ``organic'' representation of the radical anion with a localized dot in Equation~\ref{eq:birch}.
This is in contrast with the highly complex delocalized distribution of spin density over the anion observed in our simulations that follows the dynamic JT distortions of the ring.
While experimental data is currently not available for some of the properties calculated in this work, we anticipate that this will change in the future so that a more detailed understanding of the system can arise from a combination of theoretical and experimental work.

\section*{Supporting Information}

Description of computational methodology, details of data analysis, additional information on the energetics of the JT distortion of the benzene radical anion, and additional computational treatment of the calculated solute-associated vibrational spectra in Figure~\ref{fig:vib} (PDF)

Video file visualizing the pseudorotational motion of the benzene ring (MP4)

Video file visualizing the simulated trajectory of the benzene radical anion, including the evolution of the spin density (MP4)

\begin{acknowledgments}

This work was supported by the Primus16/SCI/27/247019 grant from Charles University.
K.B. acknowledges funding from the IMPRS for Many Particle Systems in Structured Environments.
This work was supported by the Project SVV 260586 of Charles University.
P.J. is thankful for support from the European Regional Development Fund (Project ChemBioDrug no. CZ.02.1.01/0.0/0.0/16\_019/0000729).
This work was supported by The Ministry of Education, Youth and Sports from the Large Infrastructures for Research, Experimental Development and Innovations Project ``IT4Innovations National Supercomputing Center -- LM2015070''.
The authors thank Christoph Schran, Tomáš Martinek, and Vojtěch Košťál for helpful comments on the manuscript.
\end{acknowledgments}

%

\end{bibunit}


\clearpage

\setcounter{section}{0}
\setcounter{equation}{0}
\setcounter{figure}{0}
\setcounter{table}{0}
\setcounter{page}{1}

\renewcommand{\thesection}{S\arabic{section}}
\renewcommand{\theequation}{S\arabic{equation}}
\renewcommand{\thefigure}{S\arabic{figure}}
\renewcommand{\thepage}{S\arabic{page}}
\renewcommand{\citenumfont}[1]{S#1}
\renewcommand{\bibnumfmt}[1]{$^{\rm{S#1}}$}

\title{Supporting information for: \mytitle}
{\maketitle}

\begin{bibunit}

\nocite{revtex-control}

\section{Computational Details}

Here, we discuss the methodology behind the AIMD simulations of benzene radical anion and of neutral benzene in bulk liquid ammonia including both the computational methods and the sequence of equilibration steps taken to obtain the production trajectories from which the results presented in the main text were extracted.

\subsection{Initial force field pre-equilibration}

To obtain structurally realistic ensemble samples to serve as initial structures for AIMD equilibration, we first approached the neutral system by an empirical force field MD (FFMD) simulation of neutral benzene in a cubic box containing 64~ammonia molecules in periodic boundary conditions.
To describe interactions, we used the custom liquid-adapted ammonia FF by Eckl et al.~\cite{Eckl2008} and a Generalized Amber FF~\cite{Wang2004} (GAFF) for benzene.
The Gromacs program~\cite{Berendsen1995} was then used for (1) assembling the system and (2) its subsequent 1.0~ns isothermal-isobaric MD equilibration at 1~bar and 223~K.
Beyond pre-equilibration purposes, this FF run was used to estimate the target volume of the simulation box for the canonical ensemble (NVT) AIMD.
This was achieved by a combination of the experimental density of liquid ammonia~\cite{Ichihara1994} with the excluded volume of benzene estimated from the simulation as experimental densities for the benzene-ammonia solution have not been reported.
Additionally, we approached the negative charge effect on the simulation box volume without performing explicit isothermal-isobaric simulations by subtracting the appropriate contribution estimated from our previous (benzene-free) equilibration simulations of solvated electrons in liquid ammonia~\cite{Buttersack2020/10.1126/science.aaz7607}.
As such, the final cubic box side length was 13.855~\AA\ for the neutral system and 13.745~\AA\ for the anion.

\subsection{AIMD methodology}

Explicit electronic structure calculations are needed for further equilibration and production simulations using AIMD.
This was realized using the CP2K program package~\cite{Hutter2014/10.1002/wcms.1159} (version 5.1) and its Quickstep module~\cite{Vandevondele2005/10.1016/j.cpc.2004.12.014,Grimme2010/10.1063/1.3382344} for efficient DFT calculations of the extended periodic bulk system within the Gaussian and plane waves (GPW) framework.
Here, we used either the generalized-gradient-approximation (GGA) revPBE~\cite{Perdew1996/10.1103/PhysRevLett.77.3865,Zhang1998/10.1103/PhysRevLett.80.890} or the hybrid revPBE0~\cite{Perdew1996/10.1103/PhysRevLett.77.3865, Zhang1998/10.1103/PhysRevLett.80.890, Adamo1999/10.1063/1.478522,Goerigk2011} density functionals (as discussed below in detail) with the D3 dispersion correction~\cite{Grimme2010/10.1063/1.3382344,Goerigk2011} added to both.
Core 1s electrons of heavy atoms (\textit{i.e.}, carbons and nitrogens) were represented by Goedecker-Tetter-Hutter pseudopotentials~\cite{Goedecker1996}.
The TZV2P basis set~\cite{Vandevondele2005/10.1016/j.cpc.2004.12.014} was used for the representation of the Kohn--Sham (KS) orbitals, while the density was represented using a plane wave basis with a 400~Ry plane-wave cutoff within the GPW method.
Additionally, for the calculations involving the hybrid functional we used the approach implemented in CP2K~\cite{Guidon2008/10.1063/1.2931945,Guidon2009/10.1021/ct900494g} with a truncated Coulomb potential with a cutoff radius of 6~\AA\ and the Auxiliary density matrix method (ADMM)~\cite{Guidon2010/10.1021/ct1002225} employing the cpFIT3 auxiliary basis set.
We also made use of the always stable predictor-corrector method~\cite{Kolafa2004} to extrapolate the density matrix times the overlap matrix to further accelerate the SCF convergence in each MD step.

The propagation of nuclei was realized with a 0.5~fs time step in the canonical ensemble using a stochastic velocity-rescaling (SVR) thermostat~\cite{Bussi2007/10.1063/1.2408420} to keep the system at the desired temperature of 223~K (ca. $-$50 $^{\circ}$C).

\subsection{AIMD Equilibration and Selection of the Density Functional}

\subsubsection*{Benzene radical anion}

Five structures evenly spaced in time were extracted from the FF simulation to serve as initial structures for AIMD equilibration.
These five structures were initially equilibrated without the excess charge (\textit{i.e.}, as neutral benzene) using the revPBE-D3 GGA functional.
The total equilibration time was 3~ps for each structure and the simulations were preformed using a Langevin thermostat with a 50~ps time constant. 
The excess electron was then introduced into the equilibrated system by changing the total charge of the system to -1.
This anionic system was then equilibrated for an additional 3~ps using revPBE-D3 and constrained DFT~\cite{Kaduk2012} (CDFT) which restricted the excess electron into the molecular subspace of the solute (defined by Becke's smoothed Voronoi partitioning~\cite{Becke1988}), allowing the surrounding solvent to adapt to the localized excess charge on the solute.

\begin{figure}[tb!]
    \centering
    \includegraphics[width=\linewidth]{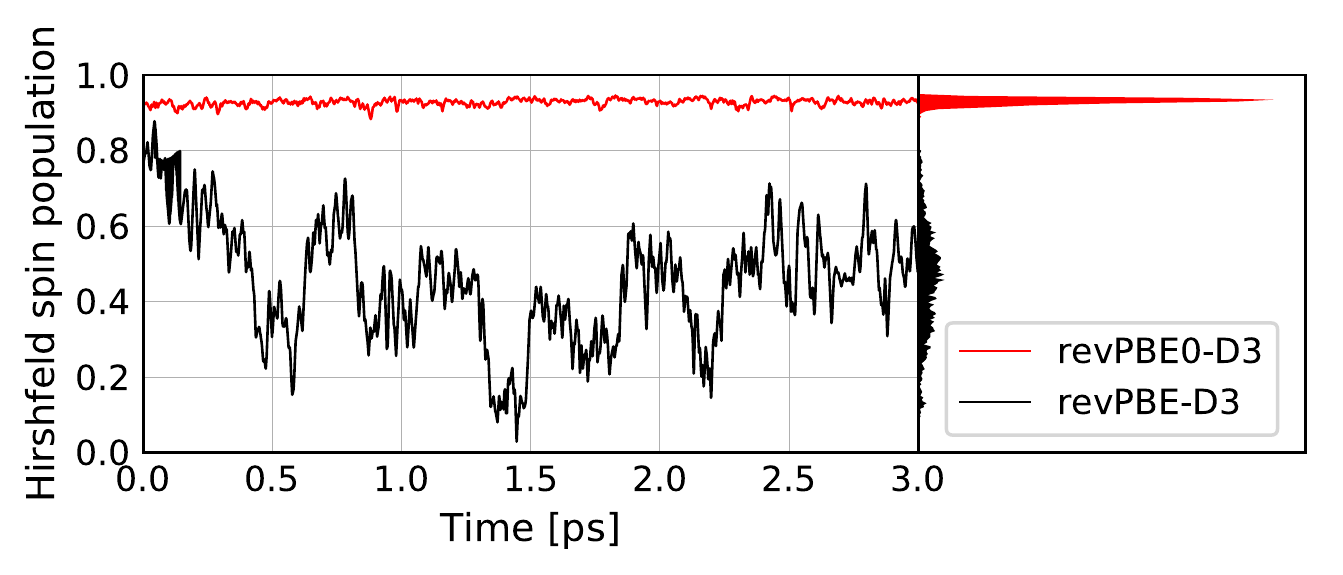}
    \caption{
    Time evolution of Hirshfeld spin population.
    The left panel shows the time evolution of the Hirshfeld spin population on the benzene radical anion over a 3~ps trajectory using the GGA revPBE-D3 (black) and hybrid revPBE0-D3 (red) functionals.
    The right panel shows the normalized distributions of the time series.
    }
    \label{fig:spin_quant}
\end{figure}

After lifting the CDFT constraint from the system, we ran unconstrained trajectories with both the GGA and the hybrid functional to check the electronic structure description and its impact on the dynamics and stability of the solute.
These sample trajectories were each 3~ps long and we calculated the Hirshfeld spin population~\cite{Hirshfeld1977} of the aromatic ring along them.
In Figure~\ref{fig:spin_quant} we plot this time evolution of the Hirshfeld spin population summed over all the 12 atoms of the aromatic ring together with their normalized distributions.

\begin{figure}[tb!]
    \centering
    \includegraphics[height=3.5cm]{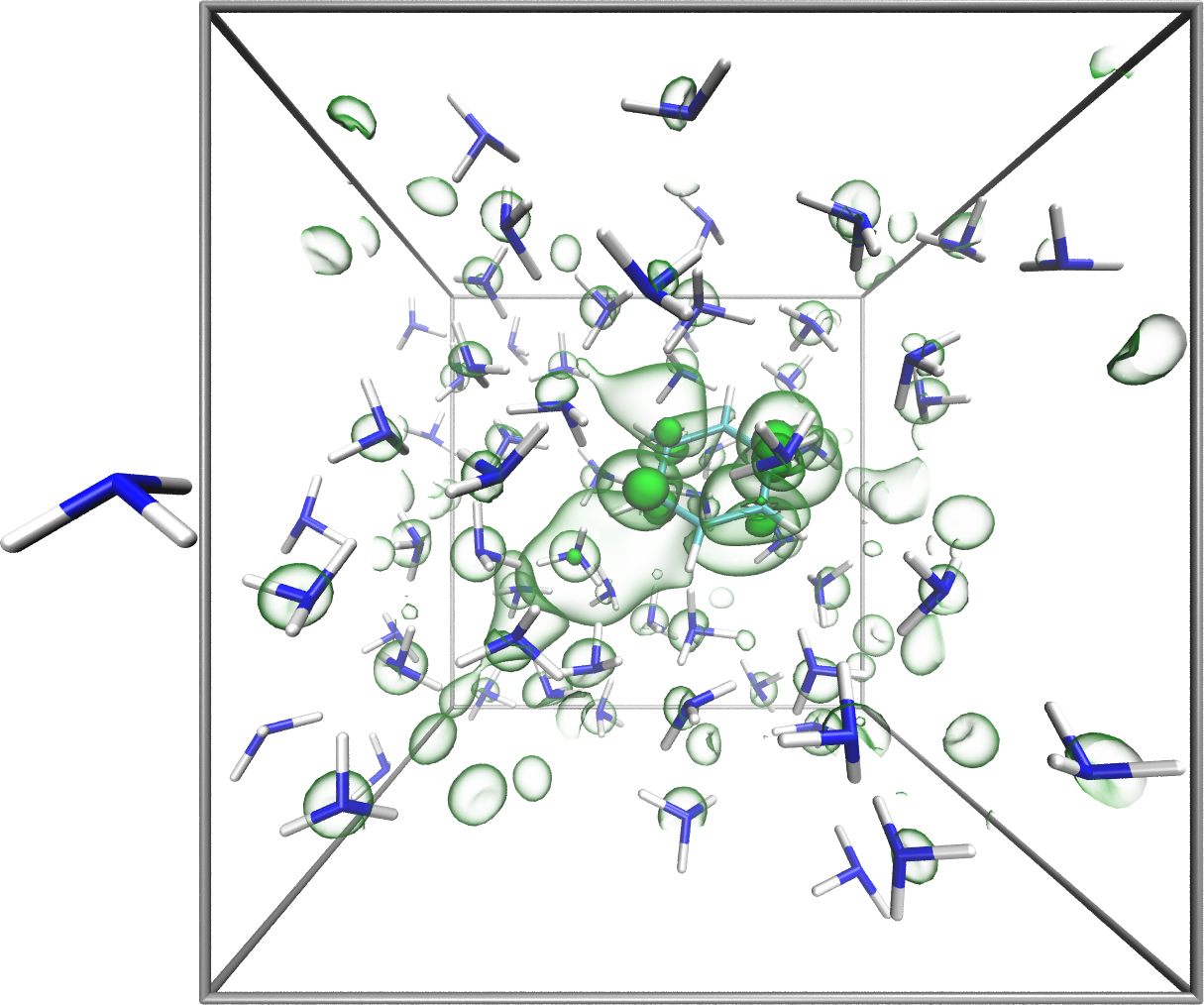}
    \includegraphics[height=3.5cm]{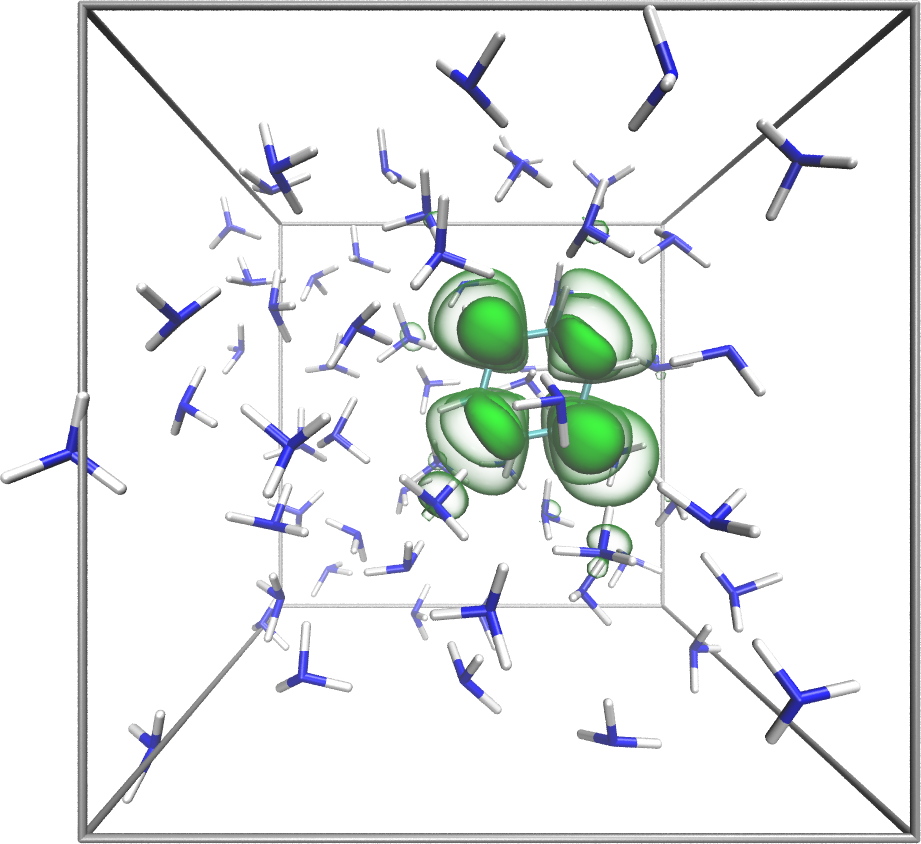}
    \caption{
    A qualitative representation of the data presented in Figure~\ref{fig:spin_quant} in the form of two simulation snapshots.
    The spin density representing the excess electron is depicted as two green contours.
    On the left, one can notice an over-delocalized spin density distribution from the GGA revPBE-D3 simulation, whereas on the right, one observes a spin density almost completely localized on the aromatic ring in the hybrid revPBE0-D3 trajectory.}
    \label{fig:spin_qual}
\end{figure}

Although both trajectories begin from the same initial structure and Kohn--Sham wave function guess (provided by the last point of the CDFT trajectory), the progress of each spin population differs instantaneously. 
The revPBE-D3 functional (black line) yields a lower initial value of the spin population and a wildly fluctuating time evolution that suggests an extensive redistribution of the spin density over the surrounding solvent.
The corresponding distribution is shallow and wide and substantially deviates from the ideal value of 1.0.
In contrast, the revPBE0-D3 hybrid functional leads to a qualitatively different result: in this case, the spin density suffers from only negligible fluctuations and the mean of its tall and narrow distribution of $\sim$0.95 is, indeed, close to unity as expected for a ``well-behaved'' anion.
This qualitative difference can be observed even at the visual level in the simulation snapshots showing the distribution of the spin density in the system.
These can be found in Figure~\ref{fig:spin_qual} where the difference between the overdelocalized distribution yielded by the revPBE-D3 functional and the neatly packed one captured in the revPBE0-D3 trajectory is obvious.
These results suggest that the spin population of the radical anion is indeed stable in  bulk liquid ammonia, albeit sensitive to methodological aspects; particularly the choice of the DFT functional.
The difference between the two functionals observed here is consistent with the well-known limitations of GGA functionals due to the self-interaction error, which is particularly pronounced for open-shell species.
This error is greatly reduced in a hybrid functional, in our case resulting in the stability described above.
Therefore, we chose the hybrid functional to simulate the main production trajectories.

\subsubsection*{Neutral benzene}

In this case, five AIMD pre-equilibrated neutral structures at the GGA level were additionally equilibrated for 3~ps at the hybrid level consistently with the equilibration of the anion.

\subsection{Production AIMD simulations}

Having decided for the methodology that reproduces the expected behavior of the radical anion, we proceeded to the production simulations.
For both the neutral and anionic systems, these were performed using the hybrid functional settings from above in the form of five independent 20~ps trajectories reaching the total simulated time of 100~ps per system.
For these production runs, the local Langevin thermostat used for equilibration was changed to a global SVR thermostat with a 200~fs time constant.

\section{Analysis Details}

Here we describe the various custom analyses which are not routinely used in computational practice and which we used to obtain the results presented in the main text.

\subsection{Procedure for angular restriction of RDFs}

As the sampling space for an unrestricted RDF is a sphere, RRDFs are obtained by sampling solvent positions from a sphere with appropriate angular sections carved out corresponding to the type of restriction to be performed.
Specifically, an axis-RRDF retains only a right circular double cone with the apex in the origin of the sphere with an inclination from the restriction axis given by the user-defined restriction angle $\omega_{\mathrm{axis}}$.
A plane-RRDF retains the complementary shape --- a sphere with a double cone removed --- with $\omega_{\mathrm{plane}}$ taken in this case from the respective plane (\textit{i.e.}, the inclination of the removed cone would be $90^\circ - \omega_{\mathrm{plane}}$).
The normalization of the RRDFs proceeds in the usual fashion as with an unrestricted RDF, however additional constant restriction-angle-dependent factors appear due to the restriction.
In particular, the volume element of a sphere
\begin{equation}
    \mathrm{d}V = 4\pi r^2 \ \dd r
\end{equation}
becomes
\begin{equation}
    \mathrm{d}V_{\mathrm{plane}} = 4\pi r^2 \sin \omega_{\mathrm{plane}} \ \dd r    
\end{equation}
and
\begin{equation}
    \mathrm{d}V_{\mathrm{axis}} = 4\pi r^2 \left( 1 - \cos \omega_{\mathrm{axis}} \right) \ \dd r 
\end{equation}
for the plane and axis restrictions, respectively.
To obtain the coordination numbers that are used to quantify the populations of the $\pi$-hydrogen sites in the main text, we formally calculate the following integral:
\begin{equation}
    N(r) = \frac{1}{2} \int_0^r 4\pi r'^2 \left(  1 - \cos\omega_{\mathrm{axis}} \right) g_{\text{axis}}(r') \ \dd r',
\end{equation}
where the $1/2$-factor ensures that the obtained population is per site, since there are two sites in total.

\subsection{Convergence of plane-RRDFs with restriction angle}

\begin{figure}[tb!]
    \centering
    \includegraphics[width=\linewidth]{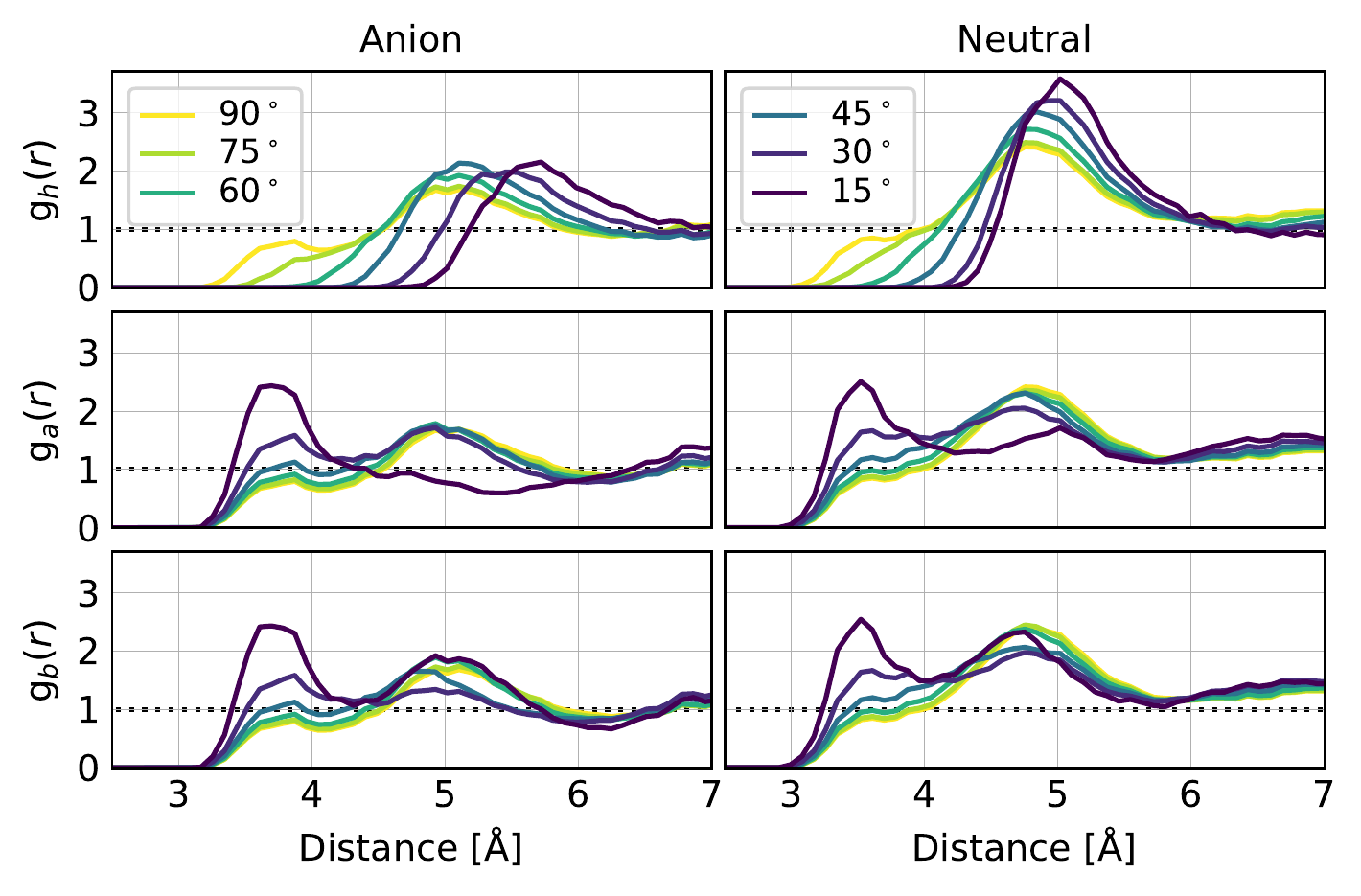}
    \caption{Restricted planar RDFs around benzene (right column) and its anion (left column) plotted at various restriction angles ranging from 90$^{\circ}$ (yellow, completely unrestricted) to 15$^{\circ}$ (purple, strongly restricted).}
    \label{fig:angle}
\end{figure}

In Figure \ref{fig:angle}, we show the convergence of the plane-RRDFs with the restriction angle ranging from 90$^{\circ}$ to 15$^{\circ}$ with 15$^{\circ}$ increments.
The restriction angle of 15$^{\circ}$ appears to be the optimal value which provides maximal definition of the individual planar contributions while not being burdened yet with statistical noise stemming from insufficient sampling.
The obtained RRDFs satisfy an important aspect of an RDF: a proper normalization is ensured through the derived additional factors as demonstrated by the correct long distance limit of 1 for all presented RRDFS.
Moreover, the RRDFs also converge smoothly to the unrestricted RDF limit at $\omega = 90^\circ$.
An identical conclusion extends to the axis-RRDF where the particular value of the restriction angle was estimated from the SDF geometry rather than from a convergence test.

\subsection{Calculation of vibrational spectra}

To obtain vibrational spectra, we exploit the Fourier transform of an autocorrelation function $c_{AA}$ of quantity $A$ to obtain its power spectrum $I_A$ through
\begin{equation}
    I_A(\omega) \propto \int_{-\infty}^{\infty} w(\tau) c_{AA}(\tau) e^{-i\omega \tau} \ \dd \tau,
\end{equation}
where $w(\tau)$ is a window function used for truncating the time series (as specified below).
To generate atomic VDOS for the system, we use the normalized velocity autocorrelation function (VACF)
\begin{equation}
    c_{\mathbf{v}\mathbf{v}}(\tau) = \sum_{i=1}^{N_\mathrm{a}} \frac{\langle \mathbf{v}_i(t_0) \cdot \mathbf{v}_i(t_0+\tau) \rangle}{\langle \mathbf{v}_i \cdot \mathbf{v}_i \rangle}
\end{equation}
defined as a sum over the $N_\mathrm{a}$ atoms of the system containing only the self terms.
As the system-wide spectrum is a straightforward sum of atomic terms, it can be split naturally into atomic or molecular contributions by limiting the sum to the subsystem in question.

In this sense, VDOS is a simpler quantity than IR absorption, where the autocorrelation function of the total dipole moment $\mathbf{M}$ is used to obtain the spectrum.
This can be decomposed into molecular correlation functions of the $N_\mathrm{m}$ molecules present in the system including the cross terms as
\begin{equation}
\begin{split}
    C_{\mathbf{MM}}(\tau) 
    & = 
    \langle
    \mathbf{M}(t_0) \cdot \mathbf{M}(t_0 + \tau)
    \rangle \\
    & =
    \sum_{i=1}^{N_\mathrm{m}}\sum_{j=1}^{N_\mathrm{m}} 
    \langle 
        \pmb{\mu}_i(t_0) \cdot \pmb{\mu}_j(t_0 + \tau) 
    \rangle.
    \label{eq:cmm}
\end{split}
\end{equation}
Again, by definition, the decomposition into spectra of individual subsystems is possible by selecting only the relevant terms from the above sum.
However, cross terms emerge alongside the self-terms in IR and must be considered in the full picture.
In this case, molecular dipole time series used to calculate IR spectra are not readily available from MD simulations such as the atomic velocity time series for VDOS are.
For the neutral benzene system, we take advantage of the CP2K option to calculate molecular dipoles on-the-fly using Maximally localized Wannier functions (MLWF) internally.
This is however implemented only for systems without spin polarization.
Thus, in the case of the spin-polarized anionic system the molecular dipoles have to be calculated manually outside CP2K.
Using MLWF centers and atomic positions, we calculate molecular dipoles for each solvent molecule in the usual fashion as a sum over all nuclei and MLWF centers associated with a given molecule multiplied by their respective charges.
This is possible because within the MLWF scheme, the excess charge is localized to the solute and all solvent molecules thus have a well-defined dipole moment independent of the point of reference.
On the other hand, the dipole of the anionic solute has to be referenced to a chosen origin.
A physically well-motivated point of reference for the solute is its COM as it moves together with the molecule and thus introduces a significant degree of translational and rotational invariance of the predicted dipole.
Using the COM reference, the solute dipole is defined as
\begin{equation}
    \pmb{\mu}_{\text{solute}} = e \sum_n \left[Z_n (\mathbf{r}_n - \mathbf{r}_{\text{COM}}) \right] - e\sum_w (\mathbf{r}_w - \mathbf{r}_{\text{COM}}),
\end{equation}
where the index $n$ runs over all nuclei of the solute, the index $w$ runs over all the MLWF centers associated with the solute, $Z_n$ is the atomic number corresponding to a given nucleus with the number of pseudopotential-represented electrons subtracted, $\mathbf{r}$ is the position vector and $e$ is the elementary charge.
To perform the actual Fourier transform on the discrete time correlation functions, we use the Fast Fourier transform (FFT) algorithm while apodizing the correlation functions by a symmetric 4000~fs Hann window and additional 10000~fs zero padding to control the noise in the resulting spectrum.

\section{Jahn--Teller distortion of the anion}

\subsection{Energetics of the Jahn--Teller distortions}

\begin{figure}[tb!]
    \centering
    \includegraphics[width=\linewidth]{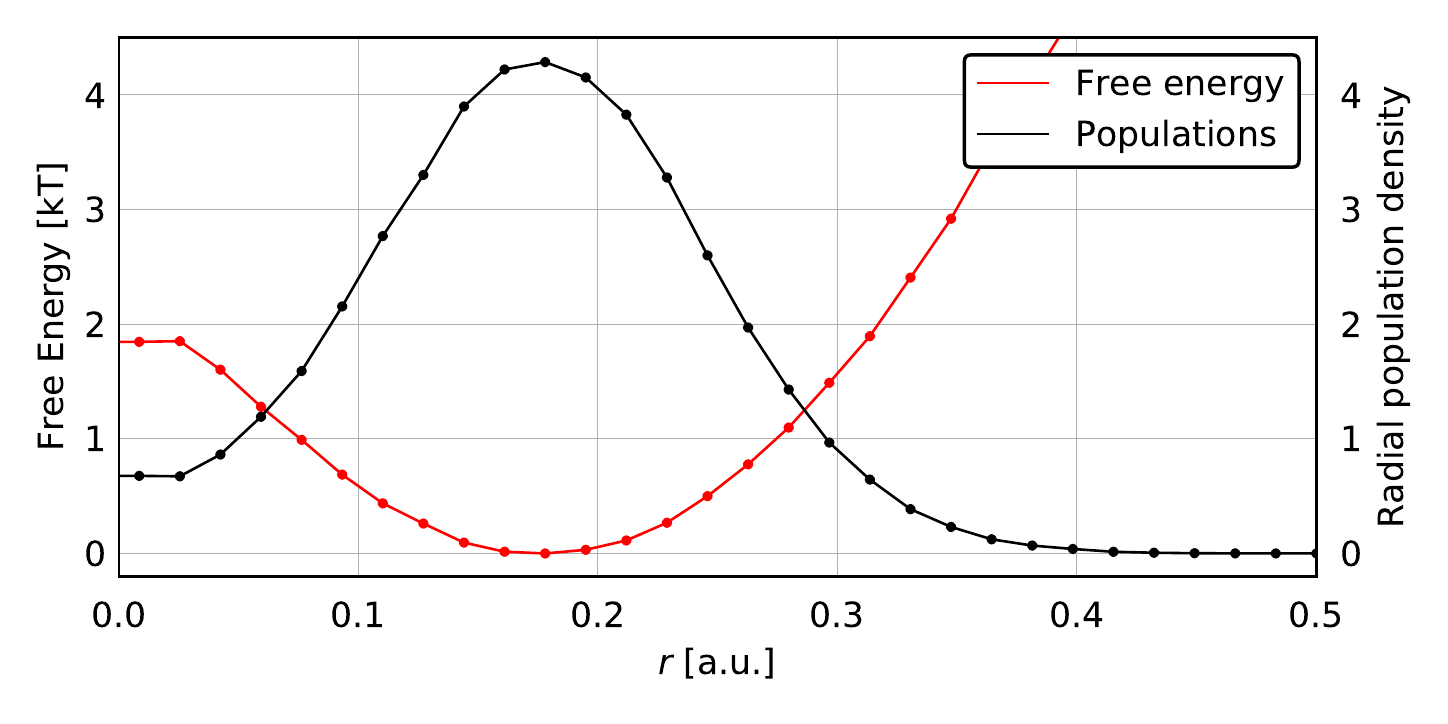}
    \caption{Populations and free energy of the JT profile in the radial direction in the space of the $Q_x$, $Q_y$ coordinates.}
    \label{fig:1D-profiles}
\end{figure}

The circular symmetry of the population profile of the JT-active modes of the anion (Figure~\ref{fig:JT2} of the main text) in principle allows to invert and integrate out one-dimensional free energy profiles in terms of polar coordinates.
The profile along the distance $r$ (Figure~\ref{fig:1D-profiles}) shows the free energy path between the point of high symmetry and the valley of distorted structures averaged over all sampled distortions.
The height of this free energy barrier is around 1~$kT$ at the temperature of the simulation, allowing the system to freely sample even the symmetric disfavored configurations with only a slight penalization by a factor of ${\sim}1/e$.
This might be an underestimate of the barrier height as the apex of the free energy profile herein might, in fact, not be the point of highest possible (\textit{i.e.}, $D_{6\mathrm{h}}$) symmetry of the molecular system, since we are averaging over the remaining vibrational modes.
In any case, the barrier depicted in Figure~\ref{fig:1D-profiles} might be considered a lower bound for the JT energy profile and the residual contributions are expected to be minor.
The complementary profile around $\theta$ addresses the free energy landscape of the pseudorotational path.
While in principle this profile should be able to energetically distinguish between the two JT structures, we are not able to obtain an energy landscape with structure above statistical noise.
This implies that the energy difference between the two structures is indeed low and the pseudorotation is effectively free.

\subsection{Time scale of Jahn--Teller distortions}

\begin{figure}[b!]
    \centering
    \includegraphics[width=\linewidth]{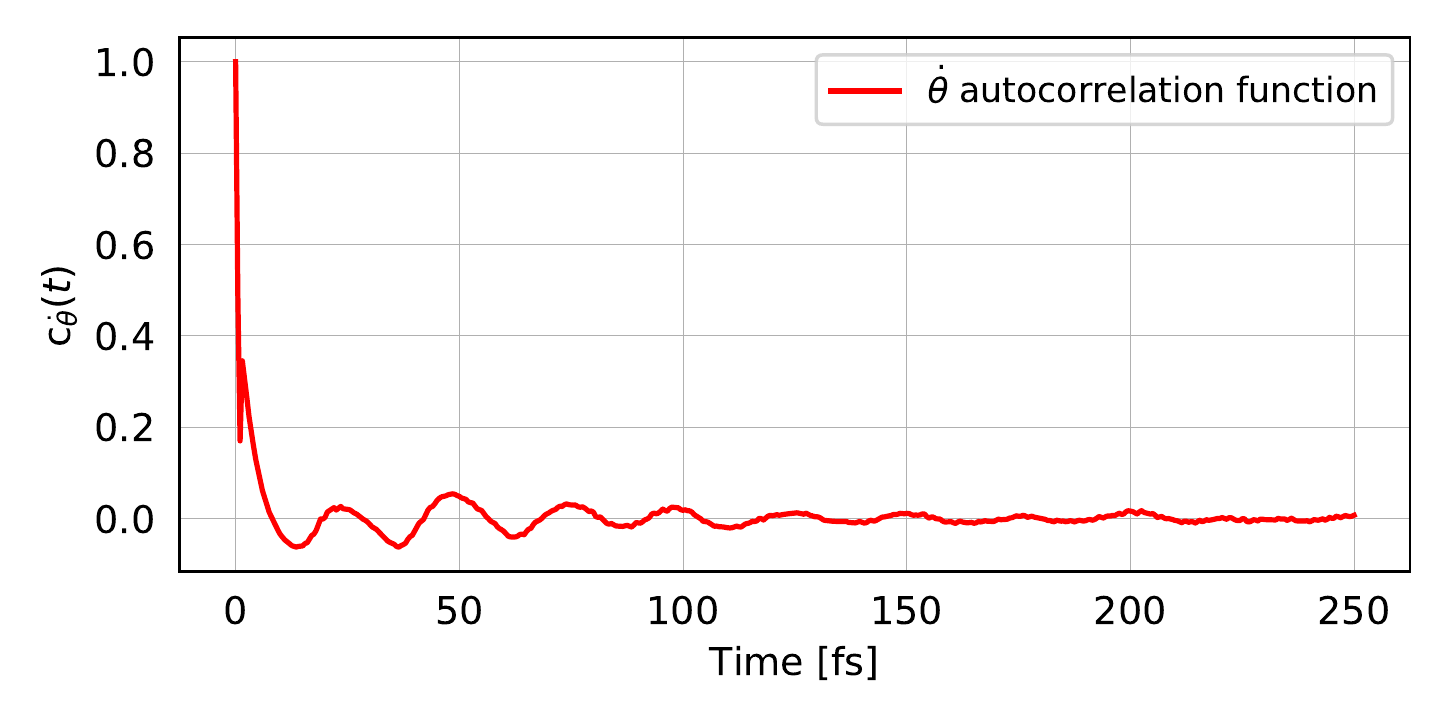}
    \caption{Normalized velocity autocorrelation function of the $\theta$ coordinate.}
    \label{fig:vacf}
\end{figure}

The ability to calculate $\theta$ from the immediate distortion of the ring allows to estimate the time scale of the JT distortion through the $\theta$ VACF
\begin{equation}
    c_{\dot{\theta}\dot{\theta}}(\tau) = \frac{\langle  \dot{\theta}(t_0) \dot{\theta}(t_0 + \tau) \rangle}{\langle \dot{\theta}^2 \rangle}.
\end{equation}
This is shown in Figure~\ref{fig:vacf} where rapid decay of the order of 10$^2$~fs can be observed.

\section{Decomposition of the IR spectra into components}

\begin{figure}[htb!]
    \centering
    \includegraphics[width=\linewidth]{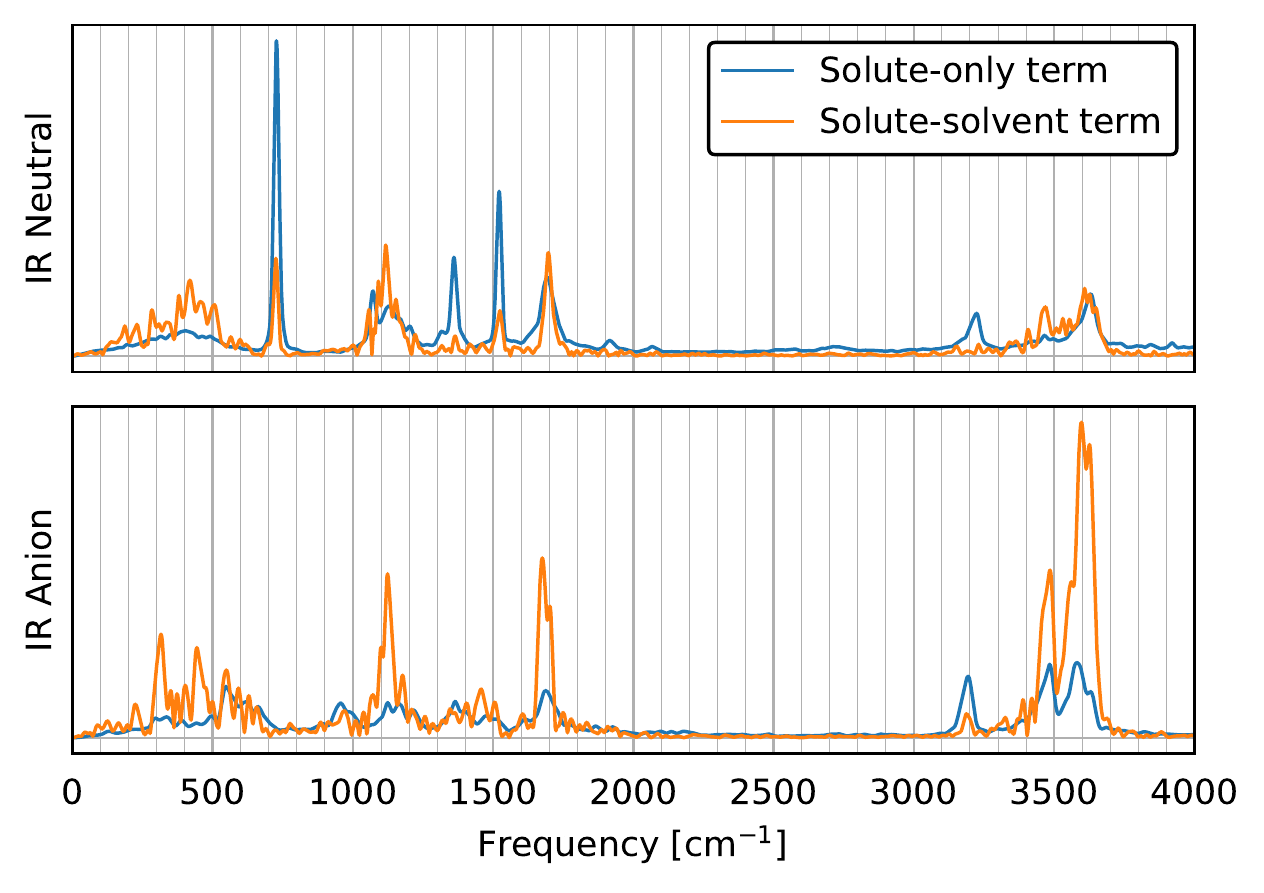}
    \caption{Benzene-ammonia solute-solvent IR spectrum (orange) compared to the solute-only spectrum (blue).}
    \label{fig:cross-term}
\end{figure}

It is natural to present solute-associated IR spectra (\textit{i.e.}, a sum of all terms in Equation~\ref{eq:cmm} involving the solute molecular dipole) in a discussion of the IR imprint of a solute.
For completeness, we show in Figure~\ref{fig:cross-term} the decomposition of the solute-associated IR spectra from the main text (Figure~\ref{fig:vib}) into the solute-only and solute-solvent contributions.
For both the neutral and anionic species, all the peaks are generally present in both components, albeit with varying intensities.

\section{$\pi$-hydrogen bonding\label{sec:vis-pi-HB}}

\begin{figure}[b!]
    \centering
    \includegraphics[width=0.5\linewidth]{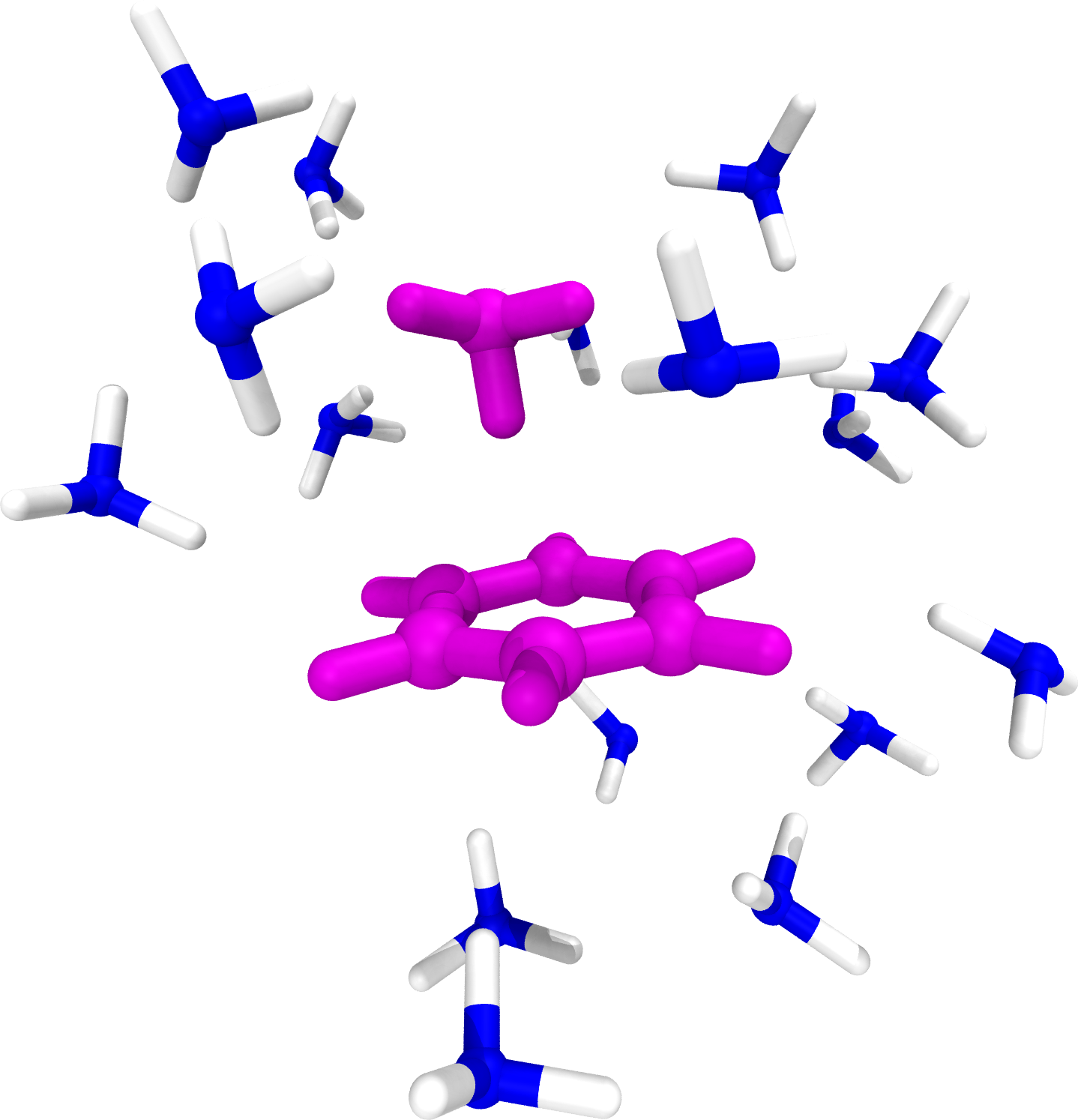}
    \caption{A simulation snapshot with the benzene radical anion solute and the $\pi$-hydrogen bonded solvent molecule in magenta and other solvent molecules within a 4~\AA\ radius in regular color coding.}
    \label{fig:pi-HB}
\end{figure}

To complement the statistical analysis on the $\pi$-hydrogen bond presented in the main text and in Figure~\ref{fig:rdfs} in the form of RRDFs and corresponding running coordination numbers, we present a snapshot visualizing the presence of a $\pi$-hydrogen bond in the radical anion trajectory in Figure~\ref{fig:pi-HB} with the interacting solvent--solute pair highlighted in magenta.

\section{Visualizations\label{sec:visualizations}}

To help with the visualization of the features of the radical anion discussed here and in the main text, we present two video files alongside this written text.

\subsection{Trajectory of the Benzene Radical Anion}

The first visualization in the file \texttt{radical-anion.mp4} is that of the actual AIMD trajectory of the benzene radical anion, including the evolution of the localized spin density with the same color coding as in Figure~\ref{fig:sim-shapshot}.
The employed time step between two neighboring frames is 2~fs, which corresponds to a stride of 4 of the original data.
With the 30 frames per second of the video file, the resulting playback speed is 60 fs/s.

\subsection{Pseudorotation}

To better relate to the nature of the pseudorotational motion of the benzene radical anion sketched out in Figure~\ref{fig:pseudorotation}, we present the video file \texttt{pseudorotation.mp4}.
This contains an artificial visualization of an optimized neutral benzene structure propagated along the two orthonormalized vibrational normal modes $Q_x$ and $Q_y$ discussed in the main text to their full extent with a $90^\circ$ phase shift and at their natural vibrational frequencies.
As this corresponds to the system circling around a unit circle in the $Q_x, Q_y$ space, one can see that for visualization purposes the distortion is amplified here by a factor of ca. 4 when compared to the physical distortion of the anion shown in Figure~\ref{fig:JT2} in the main text.
With the chosen propagation time step of 0.502~fs, the physical duration of one period of the pseudorotation of 19.09~fs, and the frame rate of 24 frames per second, the playback speed of the video is 1.58 s/period.

%

\end{bibunit}

\end{document}